\documentclass[showpacs,preprintnumbers,amsmath,amssymb,prb,aps,floatfix,twocolumn]{revtex4}
\usepackage[dvips]{graphicx}

\begin{document}

\title{Electron-lattice and strain effects in manganite heterostructures:
the case of a single interface}

\author{A. Iorio, C.A. Perroni, V. Marigliano Ramaglia and V. Cataudella}

\address{CNR-SPIN and  Dipartimento di Scienze Fisiche, \\
Universit\`a degli Studi di Napoli Federico II, \\
Complesso Universitario Monte S. Angelo, Via Cintia, I-80126 Napoli, Italy }

%\ead{vittorio.cataudella@na.infn.it}
\begin{abstract}

A correlated inhomogeneous mean-field approach is proposed in
order to study a tight-binding model of the manganite
heterostructures $(LaMnO_{3})_{2n}/(SrMnO_{3})_{n}$ with average
hole doping $x=1/3$. Phase diagrams, spectral and optical
properties of large heterostructures (up to $48$ sites along the
growth direction) with a single interface are discussed analyzing
the effects of electron-lattice anti-adiabatic fluctuations and
strain. The formation of a metallic ferromagnetic interface is
quite robust with varying the strength of electron-lattice
coupling and strain, though the size of the interface region is
strongly dependent on these interactions. The density of states
never vanishes at the chemical potential due to the formation of
the interface, but it shows a rapid suppression with increasing
the electron-lattice coupling. The in-plane and out-of-plane
optical conductivities show sharp differences since the in-plane
response has metallic features, while the out-of-plane one is
characterized by a transfer of spectral weight to high frequency.
The in-plane response mainly comes from the region between the two
insulating blocks, so that it provides a clear signature of the
formation of the metallic ferromagnetic interface.
\end{abstract}

%\pacs{75.47.Lx, 73.50.-hv, 75.10.-b}
\maketitle

\section{Introduction}
\label{Intro}

Transition metal oxides are of great current interest because of
the wide variety of the ordered phases that they exhibit and the
strong sensitivity to external perturbations. \cite{imada} Among
them, manganese oxides with formula $R_{1-x}A_{x}MnO_{3}$ ($R$
stands for a rare earth as $La$, $A$ represents a divalent alkali
element such as $Sr$ or $Ca$ and $x$ the hole doping), known as
manganites, have been studied intensively both for their very rich
phase diagram and for the phenomenon of ��colossal��
magnetoresistance. \cite{dagotto} This effect is often exhibited
in the doping regime $0.2<x<0.5$, where the ground state
of the systems is ferromagnetic. The ferromagnetic phase is
usually explained by invoking the double exchange mechanism in
which hopping of an outer-shell electron from a $Mn^{3+}$ to a
$Mn^{4+}$ site is favored by a parallel alignment of the core
spins. \cite{zener} In addition to the double-exchange term that
promotes hopping of the carriers, a strong interaction between
electrons and lattice distortions plays a non-negligible role in
these compounds giving rise to formation of polaron
quasi-particles. \cite{millis}

Very recently, high quality atomic-scale "digital"
heterostructures consisting of combination of transition metal
oxide materials have been realized. Indeed, heterostructures
represent the first steps to use correlated oxide systems in
realistic devices. Moreover, at the interface, the electronic
properties can be drastically changed in comparison with those of
the bulk. Recent examples include the formation of a thin metallic
layer at the interface between band and Mott insulators as, for
example, between $SrTiO_{3}$ ($STO$) and $LaTiO_{3}$ oxides
\cite{ohtomo} or between the band insulators \cite{ohtomo1}
$LaAlO_{3}$ and $STO$.

Very interesting examples of heterostructure are given by the
superlattices $(LaMnO_{3})_{m}/(SrMnO_{3})_{n}$ with $n/(m+n)$
average hole doping. \cite{koida} Here $LaMnO_{3}$ ($LMO$) (one
electron per $Mn$ $e_{g}$ state) and $SrMnO_{3}$ ($SMO$) (no
electrons per $Mn$ $e_{g}$ state) are the two end-member compounds
of the alloy $La_{1-x}Sr_{x}MnO_{3}$ and are both
antiferromagnetic insulating. In these systems, not only the
chemical composition but also the thickness of the constituent
blocks specified by $m$ and $n$ is important for influencing the
properties of superlattices. Focus has been on the case $m=2n$
corresponding to the average optimal hole doping $x=1/3$.
\cite{eckstein,adamo1} The superlattices exhibit a metal-insulator
transition as function of temperature for $n \leq 2$ and behave as
insulators for $n \geq 3$. The superlattices undergo a rich variety
of transitions among metal, Mott variable range hopping insulator,
interaction-induced Efros-Shklovskii insulator, and polaronic
insulator. \cite{adamo2}

Interfaces play a fundamental role in tuning the metal-insulator
transitions since they control the effective doping of the
different layers. Even when the system is globally insulating ($n
\geq 3$), some nonlinear optical measurements suggest that, for a
single interface, ferromagnetism due to double-exchange mechanism
can be induced between the two antiferromagnetic blocks.
\cite{ogawa} Moreover, it has been found that the interface density
of states exhibits a pronounced peak at the Fermi level whose
intensity correlates with the conductivity and magnetization.
\cite{eckstein1} These measurements point toward the possibility
of a two-dimensional half-metallic gas for the double-layer
\cite{ogawa1} whose properties have been studied by using
ab-initio density functional approaches. \cite{nanda} However, up
to now, this interesting two-dimensional gas has not been
experimentally assessed in a direct way by using lateral contacts on the
region between the $LMO$ and $SMO$ blocks.

In analogy with thin films, strain is another important quantity
in order to tune the properties of manganite heterostructures. For
example, far from interfaces, inside $LMO$, electron localization
and local strain favor antiferromagnetism and $e_g$ ($3z^2-r^2$)
orbital occupation. \cite{aruta} The magnetic phase in $LMO$ is
compatible with the $C$ type. \cite{dagotto} Moreover, by changing
the substrate, the ferromagnetism in the superlattice can be
stabilized. \cite{yamada}

From the theoretical point of view, in addition to $ab$ initio
calculations, tight-binding models have been used to study
manganite superlattices. Effects of magnetic and electron-lattice
interactions on the electronic properties have been investigated
going beyond adiabatic mean-field approximations.
\cite{dagotto1,millis1} However, the double layer with large
blocks of $LMO$ and $SMO$ has not been much studied. Moreover, the
effects of strain have been analyzed only within mean-field
approaches. \cite{nanda1}

In this paper we have studied phase diagrams, spectral and
optical properties for a very large bilayer $(LMO)_{2n}/(SMO)_{n}$
(up to size of $48$ planes relevant for a comparison with fabricated heterostructures)
starting from a tight binding model. We have developed a correlated inhomogeneous
mean-field approach taking into account the effects of
electron-lattice anti-adiabatic fluctuations. Strain is simulated
by modulating hopping and spin-spin interaction terms. We have
found that a metallic ferromagnetic interface forms for a large
range of the electron-lattice couplings and strain strengths. For
this regime of parameters, the interactions are able to change the
size of the interface region. We find the magnetic solutions that
are stable at low temperature in the entire superlattice. The
general structure of our solutions is characterized by three
phases running along growth $z$-direction: antiferromagnetic phase
with localized/delocalized (depending on the model parameters)
charge carriers inside $LMO$ block, ferromagnetic state at the interface
with itinerant carriers, localized polaronic $G$-type antiferromagnetic
phase inside $SMO$ block. The type of antiferromagnetic order inside
$LMO$ depends on the strain induced by the substrate.

We have discussed the spectral and optical properties
corresponding to different parameter regimes. Due to the formation
of the metallic interface, the density of states is finite at the
chemical potential. With increasing the electron-phonon
interaction, it gets reduced at the chemical potential, but it
never vanishes even in the intermediate to strong electron-phonon coupling regime.
Finally, we have studied both the in-plane and out-of-plane optical
conductivities pointing out that they are characterized by marked
differences: the former shows a metallic behavior, the latter a
transfer of spectral weight at high frequency due to the effects
of the electrostatic potential well trapping electrons in $LMO$
block. The in-plane response at low frequency is mainly due to the
region between the two insulating blocks, so that it can be used
as a tool to assess the formation of the metallic ferromagnetic
interface.

The paper is organized as follows: in sec. II the model and
variational approach are introduced, in III the results regarding the
phase diagrams are discussed, in sec. IV the spectral properties
and in sec. V the optical conductivities are analyzed, in the
final section the conclusions.

\section{The variational approach}

\subsection{Model Hamiltonian}
\label{m-va}

For manganite superlattices, the hamiltonian of the bulk $H_{0}$  has to be
supplemented by Coulomb terms representing the potential arising
from the pattern of the $La$ and $Sr$ ions, \cite{millis2} thus
\begin{equation}
H=H_{0}+H_{Coul}.
\label{ham}
\end{equation}
In order to set up an appropriate model for the double layer, it
is important to take into account the effects of the strain. The
epitaxial strain produces the tetragonal distortion of the $MnO_6$
octahedron, splitting the $e_g$ states into $x^2-y^2$ and
$3z^2-r^2$ states. \cite{nanda1} If the strain is tensile,
$x^2-y^2$ is lower in energy, while, if the strain is compressive,
$3z^2-r^2$ is favored. In the case of $n=8$ and three interfaces,
\cite{aruta} the superlattices grown on $STO$ are found to be
coherently strained: all of them are forced to the in-plane
lattice parameter of substrate and to an average out-of-plane
parameter $c \simeq 3.87 \r{A}$. \cite{aruta} As a consequence,
one can infer that $LMO$ blocks are subjected to compressive
strain $(-2.2 \%)$ and $SMO$ blocks to tensile strain $(+2.6 \%)$.
For the case of $LMO$ block, the resulting higher occupancy of
$3z^2-r^2$ enhances the out-of-plane ferromagnetic interaction
owing to the larger electron hopping out-of-plane. For the
case of $SMO$ block, the reverse occurs. A suitable model for the
bilayer has to describe the dynamics of the $e_g$ electrons which
in $LMO$ block and $SMO$ block preferentially occupy the
more anisotropic $3z^{2}-r^{2}$ orbitals and more isotropic $x^{2}-y^{2}$ orbitals,
respectively. For this reason, in this paper we adopt an effective
single orbital approximation for the bulk manganite.

The model for the bulk takes into account the double-exchange
mechanism, the coupling to the lattice distortions and the
super-exchange interaction between neighboring localized $t_{2g}$
electrons on $Mn$ ions. The coupling to longitudinal optical
phonons arises from the Jahn-Teller effect that splits the $e_g$
double degeneracy. Then, the Hamiltonian $H_{0}$ reads:

\begin{eqnarray}
H_{0}=&& - \sum_{\vec{R}_i, \vec{\delta}} t_{|\vec{\delta}|}
\left(\frac{S^{\vec{R}_i,\vec{R}_i+\vec{\delta}}_0+1/2}{2 S+1}\right) c^{\dagger}_{\vec{R}_i}c_{\vec{R}_i+\vec{\delta}}
\nonumber \\
&& +\omega_0 \sum_{\vec{R}_i}a^{\dagger}_{\vec{R}_i}a_{\vec{R}_i}
+g \omega_0 \sum_{\vec{R}_i} c^{\dagger}_{\vec{R}_i}c_{\vec{R}_i} \left( a_{\vec{R}_i}+a^{\dagger}_{\vec{R}_i}
\right)
  \nonumber \\
&& +\frac{1}{2} \sum_{\vec{R}_i,\vec{\delta}} \epsilon_{|\vec{\delta|}} \vec{S}_{\vec{R}_i} \cdot \vec{S}_{\vec{R}_i+\vec{\delta}}
- \mu \sum_{\vec{R}_i} c^{\dagger}_{\vec{R}_i} c_{\vec{R}_i} .  \label{1r}
\end{eqnarray}

Here  $t_{|\vec{\delta}|}$ is the transfer integral of electrons occupying $e_g$
orbitals between nearest neighbor ($nn$) sites,
$S^{\vec{R}_i,\vec{R}_i+\vec{\delta}}_0$ is the total spin of the subsystem
consisting of two localized spins on $nn$ sites and the conduction
electron, $\vec{S}_{\vec{R}_i}$ is the spin of the $t_{2g}$ core states
$\left( S= 3/2 \right)$, $c^{\dagger}_{\vec{R}_i} \left( c_{\vec{R}_i} \right)$
creates (destroys) an electron with spin parallel to the ionic
spin at the i-th site in the $e_g$ orbital. The coordination
vector $\vec{\delta}$ connects $nn$ sites. The first term of the
Hamiltonian describes the double-exchange mechanism in the limit
where the intra-atomic exchange integral $J$ is far larger than
the transfer integral $t_{|\vec{\delta}|}$. Furthermore, in
eq.(\ref{1r}), $\omega_0$ denotes the frequency of the local
optical phonon mode, $ a^{\dagger}_{\vec{R}_i} \left( a_{\vec{R}_i} \right)$ is
the creation (annihilation) phonon operator at the site $i$, the
dimensionless parameter $g$ indicates the strength of the
electron-phonon interaction. Finally, in Eq.(\ref{1r}), $\epsilon_{|\vec{\delta|}}$
represents the antiferromagnetic super-exchange coupling between
two $nn$ $t_{2g}$ spins and $\mu$ is the chemical potential. The
hopping of electrons is supposed to take place between the
equivalent $nn$ sites of a simple cubic lattice (with finite size
along the $z$ axis corresponding to the growth direction of the
heterostructure)  separated by the distance $|n-n^{\prime}|=a$.
The units are such that the Planck constant $\hbar=1$, the
Boltzmann constant $k_B$=1 and the lattice parameter $a$=1.

Regarding the terms due to the interfaces, one considers that
$La^{3+}$ and $Sr^{2+}$ ions act as $+1$ charges of magnitude $e$
and neutral points, respectively. In the heterostructure, the
distribution of those cations induces an interaction term for
$e_g$ electrons of $Mn$ giving rise to the Hamiltonian

\begin{eqnarray}
H_{Coul}= && \sum_{\vec{R_{i}} \neq \vec{R_{j}}}\frac{1}{2 \epsilon_d}
\frac{e^{2} n_{\vec{R}_{i}} n_{\vec{R}_{j}}}{|\vec{R_{i}}-\vec{R_{j}}|}
+\sum_{\vec{R}_{i}^{La} \neq \vec{R}_{j}^{La}}\frac{1}{2\epsilon_d}
\frac{e^{2}}{|\vec{R}_{i}^{La}-\vec{R}_{j}^{La}|}
\nonumber \\
 && -\sum_{\vec{R}_{i},\vec{R}_{j}^{La}}\frac{1}{\epsilon_d}
\frac{e^{2}n_{\vec{R}_{i}}}{|\vec{R}_{i}-\vec{R}_{j}^{La}|},
%\label{alfa}
\end{eqnarray}

with $n_{\vec{R}_{i}}=c^{\dag}_{\vec{R}_{i}} c_{\vec{R}_{i}}$ electron occupation number at $Mn$
site $i$, $\vec{R}_{i}$ and $\vec{R}_{i}^{La}$ are the positions
of $Mn$ and $La^{3+}$ in $i$th unit cell, respectively, and
$\epsilon_d$ is the dielectric constant of the material. In our
calculation the long-range Coulomb potential has been modulated by
a factor $\eta$ inducing a fictitious finite screening-length (see
Appendix A). This factor was added only for computational reasons
since it allows to calculate the summations of the Coulomb terms
over the lattice indices. We have modeled the heterostructures as
slabs whose in-plane size is infinite.

In order to describe the magnitude of the Coulomb interaction, we
define the dimensionless parameter $\alpha=e^{2}/(a\epsilon_d t_{|\vec{\delta}|})$
which controls the charge-density distribution. The order of
magnitude of $\alpha$ can be estimated from the hopping parameter
$t_{|\vec{\delta}|} \sim 0.65 eV$, lattice constant $a=4 \r{A}$, and typical value
of dielectric constant $\epsilon \sim 10$ to be around $0.2$.

Strain plays an important role also by renormalizing the
heterostructure parameters. Strain effects can be simulated by
introducing an anisotropy into the model between the in-plane
hopping amplitude $t_{\delta_{||}}=t$ (with $\delta_{||}$
indicating nearest neighbors in the $x-y$ planes) and out-of-plane
hopping amplitude $t_{|\delta_z|}=t_{z}$ (with $\delta_z$
indicating nearest neighbors along $z$ axis). \cite{dagotto2} Moreover, the strain
induced by the substrate can directly affect the patterns of core
spins. \cite{fang} Therefore, in our model, we have also considered the
anisotropy between the in-plane super-exchange energy
$\epsilon_{|\delta_{||}|}=\epsilon$ and the out-of-plane one
$\epsilon_{|\delta_z|}=\epsilon_z$. We have found that the
stability of magnetic phases in $LMO$ blocks is influenced by the
the presence of compressive strain, while in $SMO$ the sensitivity
to strain is poor. Therefore, in all the paper, we take as reference
the model parameters of the $SMO$ layers and we will consider anisotropy only
in the $LMO$ blocks with values of the ratio $t_{z}/t$ larger than unity
and of the ratio $\epsilon_z/\epsilon$ smaller than unity.

Finally, in order to investigate the effects of the
electron-lattice coupling, we will use the dimensionless quantity
$\lambda$ defined as
\begin{equation}
\lambda=\frac{g^{2} \omega_{0}}{6t}.
\end{equation}
In all the paper we will assume $\omega_{0}/t=0.5$.

\subsection{Test Hamiltonian}
In this work, we will consider solutions of the hamiltonian that
break the translational invariance in the out-of-plane
z-direction. The thickness of the slab is a parameter of the
system that will be indicated by $N_z$. We will build up a
variational procedure including these features of the
heterostructures. A simplified variational approach similar to that developed in this work
has already been proposed by some of the authors for manganite bulks \cite{perroni} and
films. \cite{perroni1,iorio}

In order to treat variationally the electron-phonon interaction,
the Hamiltonian (\ref{ham}) has been subjected to an inhomogeneous
Lang-Firsov canonical transformation. \cite{lang} It is defined by parameters
depending on plane indices along z-direction:

\begin{equation}
U=exp\left[-g\sum_{i_{||},i_{z}}(f_{i_{z}}c^{\dag}_{i_{||},i_{z}}c_{i_{||},i_{z}}+
\Delta_{i_{z}})(a_{i_{||},i_{z}}-a^{\dag}_{i_{||},i_{z}})\right],
\label{lang}
\end{equation}

where $i_{||}$ indicates the in-plane lattice sites $(i_x,i_y)$,
while $i_{z}$ the sites along the direction $z$. The quantity
$f_{i_{z}}$ represents the strength of the coupling between an
electron and the phonon displacement on the same site belonging to
$i_{z}$-plane, hence it measures the degree of the polaronic
effect. On the other hand, the parameter $\Delta_{i_{z}}$ denotes
a displacement field describing static distortions that are not
influenced by instantaneous position of the electrons.

In order to obtain an upper limit for free energy, the Bogoliubov
inequality has been adopted:

\begin{eqnarray}
F \leq F_{test}+\langle \tilde{H}-H_{test} \rangle_{t},
\label{fe}
\end{eqnarray}
where $F_{test}$ and $H_{test}$ are the free energy and the
Hamiltonian corresponding to the test model that is assumed with
an ansatz. $\tilde{H}$ stands for the transformed Hamiltonian
$\tilde{H}=UHU^{\dag}$. The symbol $\langle \rangle_{t}$ indicates
a thermodynamic average performed by using the test Hamiltonian.
The only part of $H_{test}$ which contributes to $\langle
\tilde{H}-H_{test} \rangle_{t} $ is given by the spin freedom
degrees and depends on the magnetic order of the $t_{2g}$ core
spins. For the spins, this procedure is equivalent to the standard
mean-field approach.

The model test hamiltonian, $H_{test}$, is such that that electron,
phonon and spin degrees of freedom are not interacting with each
other:

\begin{equation}
H_{test}=H^{sp}_{test}+H^{ph}_{test}+H^{el}_{test}.
\label{test}
\end{equation}

The phonon part of $H_{test}$ simply reads
\begin{eqnarray}
H^{ph}_{test}=\omega_{0}\sum_{i_{||}, i_{z}}a^{\dag}_{i_{||},i_{i_{z}}}a_{i_{||},i_{i_{z}}},
\end{eqnarray}
and the spin term is given by
\begin{equation}
H^{sp}_{test}=-g_{S}\mu_{B}\sum_{i_{||}}\sum_{i_{z}} h^{z}_{i_{||},i_{z}} S^{z}_{i_{||}, i_{z}},
\label{spin}
\end{equation}
where $g_{S}$ is the dimensionless electron-spin factor
($g_{S}\simeq 2$), $\mu_{B}$ is the Bohr magneton, and
$h^{z}_{i_{||},i_{z}}$ is the effective variational magnetic
field. In this work, we consider the following magnetic orders
modulated plane by plane:
\begin{eqnarray}
&& F,    \qquad   h^{z}_{i_{||},i_{z}}=|h^{z}_{i_{z}}|;
\nonumber \\
&& A,    \qquad   h^{z}_{i_{||},i_{z}}=(-1)^{i_z} |h^{z}_{i_{z}}|;
\nonumber \\
&& C,    \qquad   h^{z}_{i_{||},i_{z}}=(-1)^{ix+iy} |h^{z}_{i_{z}}|;
\nonumber \\
&& G,    \qquad   h^{z}_{i_{||},i_{z}}=(-1)^{ix+iy+iz} |h^{z}_{i_{z}}|.
\end{eqnarray}
For all these magnetic orders, the thermal averages of double-exchange
operator, corresponding to neighboring sites in the same plane
$i_{z}$ $\gamma_{i_{z}; i_{||},i_{||}+\delta_{||}}$ and in
different planes $\eta_{i_{z}, i_{z}+\delta_{z}; i_{||} }$,
preserve only the dependence on the $z$ plane index:
\begin{eqnarray}
\gamma_{i_{z}; i_{||},i_{||}+\delta_{||}}=\langle \frac{S^{i_{||},i_z;i_{||}+\delta_{||},i_z}_0+1/2}{2 S+1}\rangle_t=\gamma_{i_{z}}
\nonumber \\
\eta_{i_{z}, i_{z}+\delta_{z}; i_{||} }=\langle \frac{S^{i_{||},i_z;i_{||},i_z+\delta_{z}}_0+1/2}{2 S+1}\rangle_t=\eta_{i_{z}, i_{z}+\delta_{z}} .
\end{eqnarray}

In order to get the mean-field electronic Hamiltonian, we make the
Hartree approximation for the Coulomb interaction. The electronic
contribution $H^{el}_{test}$ to the test Hamiltonian becomes
\begin{eqnarray}
H^{el}_{test}&&=- t \sum_{i_{||}}\sum_{i_{z}=1}^{N_{z}}\sum_{\delta_{||}}  \gamma_{i_{z}} e^{-V_{i_{z}}}
c^{\dag}_{i_{||},i_{z}} c_{i_{||}+\delta_{||},i_{z}}
\nonumber \\
&& -t_{z} \sum_{i_{||}}\sum_{i_{z}=1}^{N_{z}}\sum_{\delta_{z}} \eta_{i_{z}, i_{z}+\delta_{z}}  e^{-W_{i_{z},i_{z}+\delta_{z}}}
c^{\dag}_{i_{||},i_{z}} c_{i_{||},i_{z}+\delta_{z}}
\nonumber \\
&& +\sum_{i_{||}}\sum_{i_{z}=1}^{N_{z}}
\left[ \phi_{eff}(i_{z})-\mu \right] c^{\dag}_{i_{||},i_z} c_{i_{||},i_z}
\nonumber \\
&& +N_{x}N_{y}(T_{1}+T_{2})+N_{x}N_{y}g^{2}\omega_{0}\sum_{i_z}\Delta_{i_z}.
\label{eltest}
\nonumber \\
\end{eqnarray}
In Eq.(\ref{eltest}), the quantity $\phi_{eff}(i_{z})$ indicates
the effective potential seen by the electrons. It consists of the
Hartee self-consistent potential $\phi(i_{z})$ (see Appendix A)
and a potential due to the electron-phonon coupling:
\begin{equation}
\phi_{eff}(i_{z})=\phi(i_{z})+ g^{2} \omega_{0} C_{i_z},
\end{equation}
with
\begin{equation}
C_{i_{z}}=f^{2}_{i_{z}}-2f_{i_{z}}+2\Delta_{i_{z}}(f_{i_{z}}-1).
\end{equation}
The factors $e^{-V_{i_{z}}}$ and $e^{-W_{i_{z},i_{z}+\delta_{z}}}$
represent the phonon thermal average of Lang-Firsov operators:
\begin{eqnarray}
e^{-V_{i_{z}}} = \langle X_{i_{||},i_z} X^{\dag}_{i_{||}+\delta_{||},i_z}\rangle_t
\nonumber \\
e^{-W_{i_{z},{i_{z}}+\delta_{z}}} = \langle X_{i_{||},i_z} X^{\dag}_{i_{||},i_z+\delta_z} \rangle_t,
\end{eqnarray}
where the operator $X_{\vec{R}_i}$ reads
\begin{equation}
X_{\vec{R}_i}= e^{g f_{i_{z}} (a_{\vec{R}_i}-a^{\dag}_{\vec{R}_i})}.
\nonumber \\
\end{equation}
Finally, the quantity $T_{1}$ and $T_{2}$ derive \label{fe} from
the Hartree approximation (see Appendix A), $N_x$ and $N_y$ denote the
size of the system along the two in-plane directions, respectively.
In order to calculate the variational free energy, we need to know eigenvalues and
eigenvectors of $H^{el}_{test}$ which depend on the magnetic order
of core spins through the double exchange terms.

\subsection{Magnetic order and diagonalization of the electronic mean-field Hamiltonian}

In order to develop the calculation, we need to fix the magnetic
order of core spins. The patters of magnetic orders is determined
by the minimization of the total free energy. By exploiting the
translational invariance along the directions perpendicular to the growth axis of the
heterostructure, the diagonalization for $H^{el}_{test}$ reduces
to an effective unidimensional problem for each pair of continuous
wave vectors $(k_{x},k_{y})=\vec{k}_{||}$. For some magnetic patterns,
the electronic problem is characterized at the interface by a
staggered structure. Therefore, we study the electron system considering a reduced
first Brillouin zone of in-plane wave vectors. To this aim, we
represent $H^{el}_{test}$ with the $2N_{z}$ states
\begin{equation}
|k_{x},k_{y},i_z\rangle,   \qquad   |k_{x}+\pi,k_{y}+\pi,i_z\rangle,
\end{equation}
with the wave vectors such that $-\pi/2 < k_{x} < \pi/2$, $-\pi/2<
k_{y}<\pi/2$, and $i_z$ going from $1$ to $N_z$. The eigenstates
of electronic test Hamiltonian are indicated by $E(k_x,k_y,n)$,
with the eigenvalue index $n$ going from $1$ to $2 N_z$. The
eigenvector related to $n$ is specified in the following way:
$b_{i_{z}}(\vec{k}_{||},n)$ for the first $N_{z}$ components,
$p_{i_{z}}(\vec{k}_{||},n)$ for the remaining $N_{z}$ components.

The variational procedure is self-consistently performed by
imposing that the total density of the system $\rho$ is given by
$N_{La}/N_{z}$, with $N_{La}$ the number of layers of $ LMO $
block, and the local plane density $\chi(i_z)$ is equal to
$\langle n_{\vec{R}_i} \rangle$. Therefore, one has to solve the
following $N_z+1$ equations:
\begin{equation}
\rho=\frac{1}{N_{x}N_{y}N_{z}}\sum_{\vec{k}_{||}}\sum_{n}n_{F} \left[ E(\vec{k}_{||},n) \right]
\end{equation}
and
\begin{eqnarray}
\chi(i_z)&=&\frac{1}{N_{x}N_{y}}\sum_{\vec{k}_{||}}\sum_{n}n_{F} \left[ E(\vec{k}_{||},n) \right]
\nonumber \\
&& \Bigg[|b_{i_{z}}(\vec{k}_{||},n)|^{2}+|p_{i_{z}}(\vec{k}_{||},n)|^{2}+
\nonumber \\
&& [b^{*}_{i_{z}}(\vec{k}_{||},n)p_{i_{z}}(\vec{k}_{||},n)+p^{*}_{i_{z}}(\vec{k}_{||},n)b_{i_{z}}(\vec{k}_{||},n)]\Bigg],
\nonumber \\
\end{eqnarray}

where $n_F(z)$ is the Fermi distribution function. These equations allow to obtain the
chemical potential $\mu$ and the the local charge density
$\chi(i_{z})$. As result of the variational analysis, one is able
to get the charge density profile corresponding to magnetic
solutions which minimize the free energy.

\section{Static properties and phase diagrams}

We have found the magnetic solutions and the corresponding density
profiles that are stable for different sizes of the $LMO$ and
$SMO$ blocks. The inhomogeneous variational approach allows to
determine the values of the electron-phonon parameters $f_{i_z}$ and
$\Delta_{i_z}$, and the magnetic order of the $t_{2g}$ spins
through the effective magnetic fields $h_{i_z}$. We will study the
systems in the intermediate to strong electron-phonon regime
characteristic of manganite materials focusing on two values of
coupling: $\lambda=0.5$ and $\lambda=0.8$. The maximum value of in-plane
antiferromagnetic super-exchange is $\epsilon=0.01 t$. The value of the Coulomb term $\alpha$ is fixed
to $\alpha=0.2$. We will analyze the heterostrucures in the low-temperature regime:
$T=0.05 t$.

The general structure of our solutions is characterized by three
phases running along $z$-direction. Actually, according to the
parameters of the model, we find $G$ or $C$ antiferromagnetic phases
corresponding to localized or delocalized charge carriers inside $LMO$ block, respectively. The
localization is ascribed to the electron-phonon coupling which
gives rise to the formation of small polarons. For the values of
$\lambda$ considered in this paper, a ferromagnetic phase always
stabilizes around the interface. The size of the ferromagnetic region
at the interface is determined by the minimization of the free energy  and
depends on the values of the system parameters. Only for larger values of
$\lambda$ and $\epsilon$, the possibility of interface
ferromagnetism is forbidden. Inside the $SMO$ block, a localized polaronic
$G$-type antiferromagnet phase is always stable.

\begin{figure}
\begin{center}
%\hspace{-3.65cm}
\includegraphics[width=0.5\textwidth, angle=0]{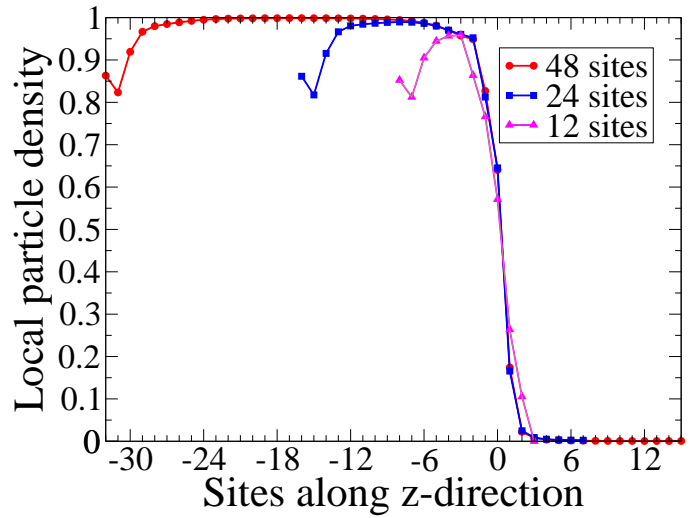}%{phase_3d_scaf.png}
\end{center}
\caption{Comparison among density profiles corresponding to different sizes at $\lambda=0.5$ and
$\epsilon=0.01t$. The index $0$ indicates the interface $Mn$-plane between the last $La$-plane in $LMO$
block and the first $Sr$-plane in $SMO$ block.}
\label{f1}
\end{figure}

At first, we have analyzed the scaling of the static properties as
function of the size of the system along the $z$ growth direction.
Therefore, a comparison of the density profiles has been done with $(LMO)_{8}/(SMO)_{4}$,
$(LMO)_{16}/(SMO)_{8}$ and $(LMO)_{32}/(SMO)_{16}$ systems. In
Fig. \ref{f1}, we show the density profiles in a situation where
strain-induced anisotropy has not been introduced. It is
worth noticing that we indicate the interface  $Mn$-plane between
the last $La$-plane in $LMO$ block and the first $Sr$-plane in
$SMO$ block with the index $0$. For a sufficiently large numbers
of planes, the charge profile along $z$ shows a well-defined
shape. Indeed, the local density is nearly unity in $LMO$ block,
nearly zero in $SMO$ block, and it decreases from $1$ to $0$ in the
interface region. The decrease of charge density for the first
planes of $LMO$ is due to the effect of open boundary conditions
along the $z$ direction. In the intermediate electron-phonon
coupling regime that we consider in Fig. (\ref{f1}), the region
with charge dropping involves $4-5$ planes between the two blocks.
We notice that the local charge density for $(LMO)_{16}/(SMO)_{8}$
and $(LMO)_{32}/(SMO)_{16}$ systems are very similar around the
interface. Furthermore, the numerical results show close values of
variational free energy corresponding to above mentioned systems.
Given the similarity of the properties of these two systems, in
the following, we will develop the analysis on the role of
interface studying the system $(LMO)_{16}/(SMO)_{8}$.

For the same set of electron-phonon and magnetic couplings, the
variational parameters and the Hartree self-consistent potential
along z-axis are shown in Fig. 2. The effective magnetic fields
are plotted for the most stable magnetic solution: antiferro
$G$ orders well inside $LMO$ (planes $1-15$) and $SMO$ (planes $19-24$), and 
ferromagnetic planes at the interface (planes $16-18$). 
The peak in the plot of the magnetic fields signals that ferromagnetism
is quite robust at the interface. The variational electron-phonon 
parameters $f_{i_z}$ are small on the $LMO$ side and at the interface, 
but close to unity in $SMO$ block. This means that, for these values of the couplings,
carriers are delocalized in $LMO$ up to the interface region, but
small polarons are present in the $SMO$ block. The quantities
$\Delta_{i_z}$, entering the variational treatment of the
electron-phonon coupling, are determined by $f_{i_z}$ and the
local density $<n_{i_z}>$ through the equation:
$\Delta_{i_z}=<n_{i_z}>(1-f_{i_z})$. The Hartree self-consistent
potential $\Phi$ indicates that charges are trapped into a potential well
corresponding to the $LMO$ block. Moreover, it is important to
stress the energy scales involved in the well: the barrier between
$LMO$ and $SMO$ block is of the order of the electron band-width.
Furthermore, at the interface, the energy difference between
neighboring planes is of the order of the hopping energy $t$.

\begin{figure}
%\hspace{-3.65cm}
\includegraphics[width=0.5\textwidth,height=0.30\textheight,angle=0]{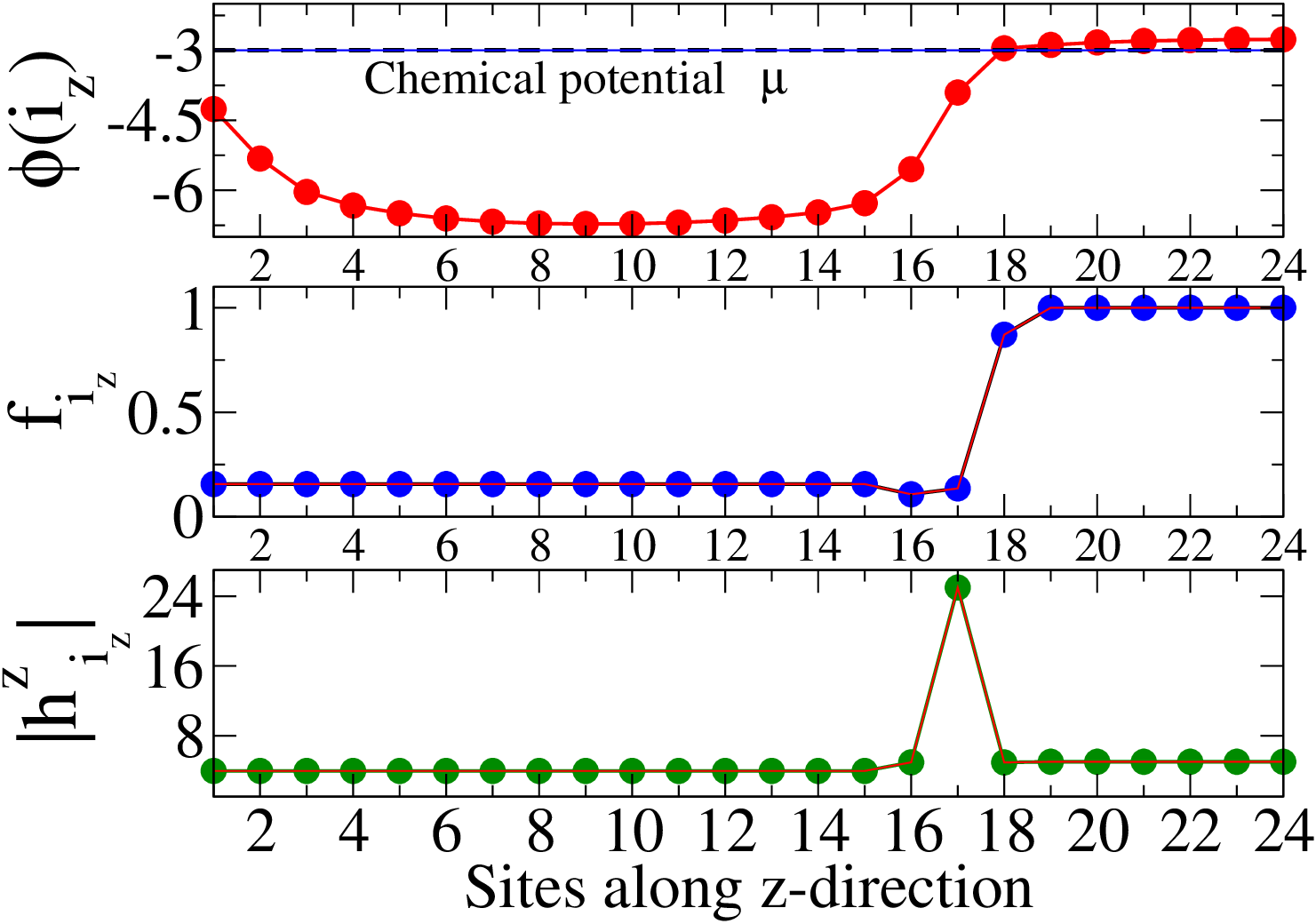}%{phase_3d_scaf.png}
\caption{Self-consistent Hartree potential $\phi(i_{z})$ (upper panel, in units of $t$), variational parameters $f_{i_z}$ (mid panel)
and effective magnetic fields $|h^{z}_{i_{z}}|$ (lower panel) along the z-axis for $\lambda=0.5$ and
$\epsilon=0.01t$.}
\label{f2}
\end{figure}

As mentioned above, for these systems, strain plays
an important role. In order to study quantitatively its effect,
we have investigated the phase diagram under the variation of the
hopping anisotropy $t_{z}/t$ for two different values of $\epsilon_{z}$ ($\epsilon_z = \epsilon =0.01 t$,
$\epsilon_z = 0$). Indeed, we simulate the compressive strain in the $LMO$ block
increasing the ratio $t_{z}/t$ and decreasing $\epsilon_z / \epsilon$. On the other hand, the tensile
strain in the $SMO$ block favour the more isotropic $x^2-y^2$ orbital and does not
yield sizable effects. Therefore, for the $SMO$ block, in the following, we choose
$t_{z}=t$ and $\epsilon_z = \epsilon$. For what concerns the electron-phonon interaction,
we assume an intermediate coupling, $\lambda=0.8$.
As shown in the upper panel of Fig. 3, with increasing the ratio $t_{z}/t$ up to $1.7$ for $\epsilon_z = \epsilon$,
the magnetic order in $LMO$ does not change since it remains $G$
antiferromagnetic. However, the character of charge carriers is
varied. Actually, for $\lambda=0.8$, in the absence of anisotropy,
small polarons are present in the $LMO$ block.
Moreover, at $t_{z}/t \simeq 1.5$, in $LMO$, a change from
small localized polarons to large delocalized polaron occurs. For all values of the ratio $t_{z}/t$,
the interface region is characterized by ferromagnetic order with large polaron carriers and
$SMO$ by $G$ antiferromangnetic order with small polaron carriers.

\begin{figure}
%\hspace{-3.65cm}
\includegraphics[width=0.5\textwidth,height=0.30\textheight,angle=0]{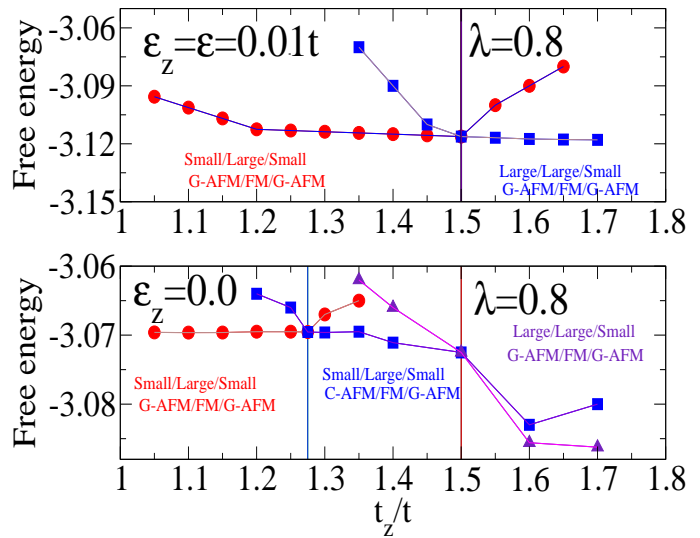}
\caption{Phase diagram in the hopping anisotropy-energy plane for $LMO_{16} SMO_{8}$ system,
corresponding to $\lambda=0.8$ for $\epsilon_{z}=0.01t$ (upper panel) and $\epsilon_{z}=0$ (lower panel).}
%A\B\C refers to magnetic orders and character of charge carriers inside $LMO$ %$(A)$, at interface $(B)$, inside $SMO$ $(C)$.}
\label{f4}
\end{figure}

It has been shown that it is also important to consider the anisotropy in super-exchange
($\epsilon_{z} \neq \epsilon$) parameters as consequence of strain. \cite{fang}
In order to simulate the effect of compressive strain in $LMO$,
a reduction of $\epsilon_{z}$ will be considered. We discuss
the limiting case: $\epsilon_{z}=0$. For this regime of parameters,
the effect on the magnetic phases is the strongest. As shown in the lower panel of Fig. 3,
for $1.28 \le t_{z}/t \le 1.5$, in $LMO$ block, a $C$-type antiferromagnetic
phase is the most favorable. The transition from small to large
polaron again takes place at $t_{z}/t \simeq 1.5$. Therefore,
we have shown that there is a range of parameters where $LMO$ block has
$C$-type antiferromagnetic order with small localized polarons. Due to the effect of strain,
the magnetic solution in $LMO$ turns out to be compatible with experimental
results in superlattices. \cite{aruta}
The interface is still ferromagnetic with metallic large polaron
features. In the figure $A$/$B$/$C$ refers to magnetic orders and character of charge carriers inside $LMO$ (A),
at interface (B), inside $SMO$ (C).

In order to analyze the effects of the electron-phonon interaction, a comparison
between two different electron-phonon couplings is reported in Fig. 4. We have investigated the
solutions which minimize the variational free energy at fixed
value of the anisotropy factors $t_{z}/t=1.3$ and $\epsilon_z=0$
at $\lambda=0.5$ and $\lambda=0.8$. The
magnetic solution in $LMO$ block is $C$ antiferromagnetic until
the $15th$ plane. For both values of $\lambda$, polarons are
small. In $SMO$ block, starting from the $19th$ plane, the
solution is $G$-type antiferromagnetic together with localized
polarons. Three planes around the interface are ferromagnetically
ordered. For $\lambda=0.5$, all the three planes at the interface
are characterized by delocalized polarons, while, for
$\lambda=0.8$, only the plane linking the ends of $LMO$ and $SMO$
blocks is with delocalized charge carriers.

As shown in Fig. 4, the quantity $\lambda$ has important
consequences on the physical properties such as the local particle
density. Actually, for $\lambda=0.8$ the transition from occupied
to empty planes is sharper at the interface. Only one plane at
the interface shows an intermediate density close to $0.5$. For
$\lambda=0.5$ the charge profile is smoother and the three
ferromagnetic planes with large polarons have densities different
from zero and one.

For the analysis of the spectral and optical
quantities, we will consider the parameters used for the
discussion of the results in this last figure.

\begin{figure}
%\hspace{-3.65cm}
\includegraphics[width=0.5\textwidth, angle=0]{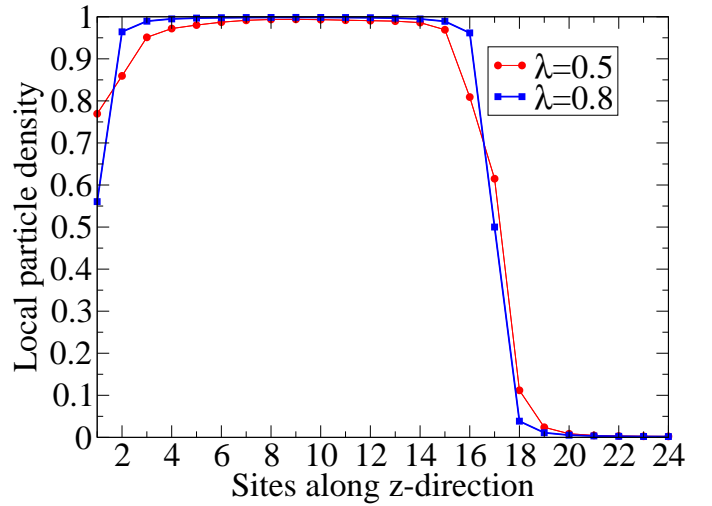}%{phase_3d_scaf.png}
\caption{Comparison between local particle density corresponding to $\lambda=0.5$ and
$\lambda=0.8$.}
\label{f5}
\end{figure}

\section{Spectral properties}

In the following section we will calculate the spectral properties of the
heterostructure for the same parameters used in Fig. 4.

Performing the canonical transformation (\ref{lang}) and exploiting the
cyclic properties of the trace, the electron Matsubara Green's function
becomes
\begin{eqnarray}
%\mathcal{G}(\tao)_{i_{x},i_{y},i_{z};i_{x},i_{y},i_{z}}=-\langle(c_{i_{x},i_{y},i_{z}}c^{\dag}_{i_{x},i_{y},i_{z}}\rangle
\mathcal{G}(\vec{R}_{i},\vec{R}_{j},\tau)=-\langle T_{\tau}c_{\vec{R}_i}(\tau)X_{\vec{R}_i}(\tau)c^{\dag}_{\vec{R}_j}(0)X^{\dag}_{\vec{R}_j}(0)\rangle.
\end{eqnarray}
By using the test Hamiltonian (\ref{test}), the correlation
function can be disentangled into electronic and phononic terms. \cite{perroni,perroni1}
Going to Matsubara frequencies and making the analytic
continuation $i\omega_{n} \rightarrow \omega+i\delta$, one obtains
the retarded Green's function and the diagonal spectral function
$A^{i_{x}i_{y}}_{i_{z}}(\omega)$ corresponding to
$\vec{R}_{i}=\vec{R}_{j}$
\begin{eqnarray}
&& A^{i_{x},i_{y}}_{i_{z}}(\omega)=
\nonumber \\
&& e^{S_T^{i_{z}}}\sum_{l=- \infty}^{\infty} I_{l} (S^{i_{z}})e^{\frac{\beta l\omega_{0}}{2}}[1-n_{F}(\omega-l\omega_{0})]
g^{i_{x},i_{y}}_{i_{z}}(\omega-l\omega_{0})
\nonumber \\
&& +e^{S_T^{i_{z}}}\sum_{l=-\infty}^{\infty} I_{l} (S^{i_{z}})e^{\frac{\beta l\omega_{0}}{2}} n_{F}(\omega+l\omega_{0})
g^{i_{x},i_{y}}_{i_{z}}(\omega+l \omega_{0}),
\nonumber \\
\label{A}
\end{eqnarray}
where $S^{i_{z}}_{T}=g^{2}f^{2}_{i_{z}}(2N_{0}+1)$, $S^{i_{z}}=2g^{2}f^2_{i_{z}}[N_0(N_{0}+1)]^{\frac{1}{2}}$, $I_l(z)$ modified
Bessel functions,
and $g^{i_{x},i_{y}}_{i_{z}}(\omega)$ is
\begin{eqnarray}
&& g^{i_{x},i_{y}}_{i_{z}}(\omega)=\frac{2 \pi}{N_{x}N_{y}}\sum_{\vec{k}_{||}}\sum_{n=1}^{2N_{z}}\delta[\omega-E(\vec{k}_{||},n)]
\nonumber \\
&& \times \Bigg[|b_{i_{z}}(\vec{k}_{||},n)|^{2}+|p_{i_{z}}(\vec{k}_{||},n)|^{2}+ \nonumber \\
&& (-1)^{i_{x}+i_{y}}[b^{*}_{i_{z}}(\vec{k}_{||},n)p_{i_{z}}(\vec{k}_{||},n)
+p^{*}_{i_{z}}(\vec{k}_{||},n)b_{ic_{z}}(\vec{k}_{||},n)]\Bigg].
\nonumber \\
\end{eqnarray}
The density of states $D(\omega)$ is defined as
\begin{equation}
D(\omega)=\frac{1}{N_{x}N_{y}N_z} \frac{1}{2 \pi} \sum_{i_x,i_y,i_z}   A^{i_{x},i_{y}}_{i_{z}}(\omega). \end{equation}

\begin{figure}
%\hspace{-3.65cm}
\includegraphics[width=0.5\textwidth, height=0.25\textheight, angle=0]{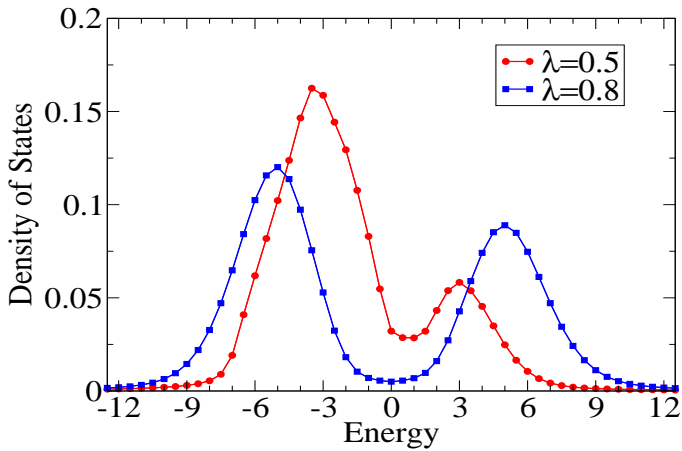}%{phase_3d_scaf.png}
\caption{Comparison between density of states (in units of $1/t$) as a function of the
energy (in units of $t$) corresponding to $\lambda=0.5$ and $\lambda=0.8$.}
\label{f6}
\end{figure}

In Fig. 5 we report the density of state of the system
$(LMO)_{16}/(SMO)_{8}$. It has been calculated measuring
the energy to the chemical potential $\mu$. This
comparison has been made at fixed low temperature
($K_{B}T=0.05t$), therefore we can consider the chemical potential
very close to the Fermi energy of the system. At $\lambda=0.5$,
the spectral function exhibits a residual spectral weight at $\mu$. The main
contribution to the density of states at the chemical potential $\mu$ comes from the three
ferromagnetic large polaron planes at the interface. Indeed, the
contributions due to the ($LMO$) and ($SMO$) blocks is negligible.

For stronger electron-phonon coupling at $\lambda=0.8$, we observe an important depression of
the spectral function at $\mu$. Hence the formation of a clear
pseudogap takes place. This result is still compatible with the
solution of our variational calculation since, for this value of
$\lambda=0.8$, there is only one plane with delocalized charge carriers
which corresponds to the plane indicated as the interface
($i_{z}=17$), while the two further ferromagnetic planes around
the interface are characterized by small polarons. The depression
of the density of the states at the Fermi energy is due also to the
polaronic localization well inside the $LMO$ and $SMO$ block.
In any case we find that, even for  $\lambda=0.8$, the density of states never vanishes
at the interface in agreement with experimental results. \cite{eckstein1}

In this section we have found strong indications that a metallic ferromagnetic
interface can form at the interface between $LMO$ and $SMO$ blocks. This situation
should be relevant for superlattices with $n \geq 3$, where resistivity measurements
made with contacts on top of $LMO$ show a globally insulating behavior. 
In our analysis we have completely neglected any effect due to disorder even if, 
both from experiments \cite{eckstein,adamo1} and theories \cite{dagotto1}, it has been suggested that localization
induced by disorder could be the cause of the metal-insulator
transition observed for $n \geq 3$. We point out that
the sizable source of disorder due to the random doping with $Sr^{2+}$ is
strongly reduced since, in superlattices, $La^{3+}$ and $Sr^{2+}$
ions are spatially separated by interfaces. Therefore, the amount
of disorder present in the heterostructure is strongly reduced in comparison with the alloy.
However, considering the behavior of the $LMO$ ($SMO$) block
as that of a bulk with a small amount of holes (particles), one expects that even a
weak disorder induces localization. On the other hand, a weak disorder is not able 
to prevent the formation of the ferromagnetic metallic interface favored by
the double-exchange mechanism and the charge transfer between the bulk-like blocks:
the states at the Fermi level due to the interface formation have enough density \cite{eckstein1} so that
they cannot be easily localized by weak disorder. In this section, we have shown that this can be the case
in the intermediate electron-phonon coupling regime appropriate for $LMO/SMO$ heterostructures.

In the next section we will analyze the effects of electron-phonon coupling and strain
on the optical conductivity in the same regime of the parameters considered in this section.

\section{Optical properties}

To determine the linear response to an external field of frequency
$\omega$, we derive the conductivity tensor
$\sigma_{\alpha,\beta}$ by means of the Kubo formula. In order to
calculate the absorption, we need only the real part of the
conductivity
\begin{eqnarray}
Re \sigma_{\alpha,\alpha}(\omega)=-\frac{Im \Pi^{ret}_{\alpha,\alpha}}{\omega},
\label{realsig}
\end{eqnarray}
where $\Pi^{ret}_{\alpha,\beta}$ is the retarded current-current
correlation function. Following a well defined scheme \cite{perroni,perroni1}
and neglecting vertex corrections, one can get a compact
expression for the real part of the conductivity
$\sigma_{\alpha,\alpha}$. It is possible to get the conductivity
both along the plane perpendicular to growth axis, $\sigma_{xx}$,
and parallel to it, $\sigma_{zz}$. In order to calculate the
current-current correlation function, one can use the spectral
function $A_{\vec{k}_{||};i_{z},j_{z}}$ derived in the previous
section exploiting the translational invariance along in-plane
direction. It is possible to show that the components of the real
part of the conductivity become
\begin{eqnarray}
Re[\sigma_{xx}](\omega)=\frac{e^{2}t^{2}}{N_{x}N_{y}}\sum_{k_{x},k_{y}}4sen^2(k_{x})
\frac{1}{N_{z}}\sum_{i_{z},j_{z}}\gamma_{i_{z}}\gamma_{j_{z}}
\nonumber \\
\times \frac{1}{\omega}\int^{\infty}_{-\infty}\frac{d\omega_{1}}{4\pi}[n_{F}(\omega_{1}-\omega)-n_{F}(\omega_{1})]
\nonumber \\
\times A_{k_{x},k_{y};i_{z},j_{z}}(\omega_{1}-\omega)A_{k_{x},k_{y};i_{z},j_{z}}(\omega_{1}), \end{eqnarray}
and
\begin{eqnarray}
&& Re [\sigma_{zz}](\omega)=\frac{e^2t^2}{N_xN_y}\sum_{k_x,k_y}\frac{1}{N_z}\sum_{i_z,j_z} \sum_{\delta_{1z},\delta_{2z}}
\delta_{1z} \delta_{2z}
\nonumber \\
&&\times  \eta_{i_z,i_z+\delta_{1z}} \eta_{j_z,j_z+\delta_{2z}} \frac{1}{\omega}\int^{\infty}_{-\infty}\frac{d\omega_1}{4\pi}[n_F(\omega_1-\omega)-n_F(\omega_1)]
\nonumber \\
&&\times A_{k_x,k_y;i_z+\delta_{1z},j_z+\delta_{2z}}(\omega_1-\omega)A_{k_x,k_y;i_z,j_z}(\omega_1).
\end{eqnarray}

\begin{figure}
%\hspace{-3.65cm}
\includegraphics[width=0.55\textwidth, angle=0]{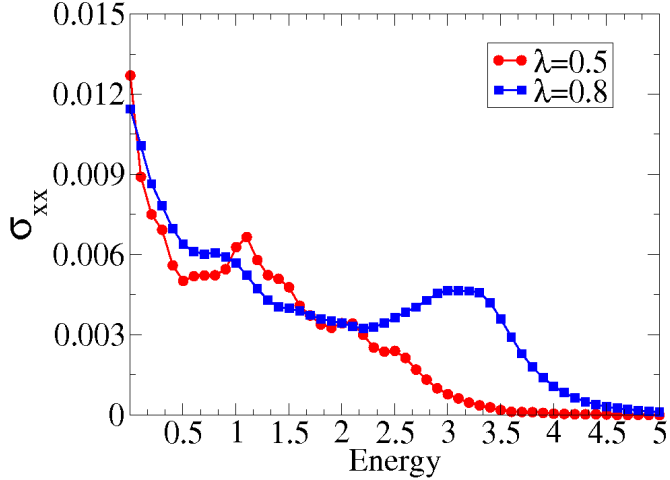}%{phase_3d_scaf.png}
\caption{The conductivity (in units of $e^2/(mt)$, with $m=1/(2t)$) into the plane perpendicular to
growth direction of the $(LMO)_{16}/(SMO)_{8}$ bilayer as a function of the energy  (in units of $t$)
for different values of $\lambda$.}
\label{f6}
\end{figure}

In Fig. 6, we report the in-plane conductivity as function of the
frequency at $\lambda=0.5$ and $\lambda=0.8$. We have checked that
the in-plane response mainly comes from the interface planes. Both
conductivities are characterized by a Drude-like response at low
frequency. Therefore, the in-plane conductivity provides a clear
signature of the formation of the metallic ferromagnetic
interface. However, due to the effect of the interactions, we have found that the low
frequency in-plane response is at least one order of magnitude
smaller than that of free electrons in the heterostructures.
Moreover, additional structures are present in the absorption with
increasing energy. For $\lambda=0.5$, a new band with a peak
energy of the order of hopping $t=2 \omega_0$ is clear in the
spectra. This structure can be surely ascribed to the presence of
large polarons at the three interface planes. \cite{perroni} Actually, this band
comes from the incoherent multiphonon absorption of large
polarons at the interface. This is also confirmed by the fact that this band is
quite broad, therefore it can be interpreted in terms of
multiple excitations. For $\lambda=0.8$, the band is even
larger and shifted at higher energies. In this case, at the
interface, large and small polarons are present with a
ferromagnetic spin order. Therefore, there is a mixing of
excitations whose net effect is the transfer of spectral weight at
higher frequencies.

The out-of-plane optical conductivities show significant differences in
comparison with the in-plane responses. In Fig. 7, we report
out-of-plane conductivity as function of the frequency at
$\lambda=0.5$ and $\lambda=0.8$. First, we observe the absence of
the Drude term. Moreover, the band at energy about $2 \omega_0$ is narrower
than that in the in-plane response. Therefore, the origin of this
band has to be different. Actually, the out-of-plane optical
conductivities are sensitive to the interface region. A charge
carrier at the interface has to overcome an energy barrier in order
to hop to the neighbour empty site. As shown in Fig. 2, the typical energy
for close planes at the interface is of the order of the hopping $t$. Therefore,
when one electrons hops along $z$, it has to pay at least an energy of the
order of $t$. In the out-of-plane spectra, the peaks at low energy
can be ascribed to this process. Of course, by paying a larger
energy, the electron can hop to next nearest neighbors. This
explains the width of this band due to inter-plane hopping.

Additional structures are present at higher energies in the
out-of-plane conductivities. For $\lambda=0.5$ the band at high
energy is broad with small spectral weight. For $\lambda=0.8$,
there is an actual transfer of spectral weight at higher energies.
A clear band is peaked around $10 t$. This energy scale can be
intepreted as given by $2 g^2 \omega_0=9.6 t$ for $\lambda=0.8$.
Therefore, in the out-of-plane response, the contribution at high energy can
be interpreted as due to small polarons. \cite{perroni,mahan}

\begin{figure}
%\hspace{-3.65cm}
\includegraphics[width=0.5\textwidth, angle=0]{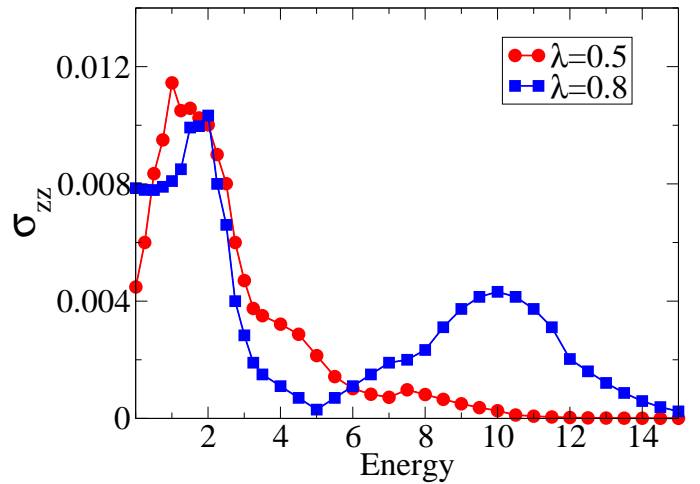}%{phase_3d_scaf.png}
\caption{The conductivity (in units of $e^2/(mt)$, with $m=1/(2t)$) along the growth direction of the $(LMO)_{16}(SMO)_{8}$
bilayer as a function of the energy (in units of $t$) for $\lambda=0.5$ and $\lambda=0.8$.}
\label{f6}
\end{figure}

Unfortunately, experimental data about optical properties of the $LMO/SMO$ bilayers are
still not available. Therefore, comparison with experiments is not possible.
Predictions about the different behaviors among $\sigma_{xx}$ and $\sigma_{zz}$ can be easily checked
if one uses in-plane and out-of-plane polarization of the electrical fields used in the experimental probes. More important,
the formation of two-dimensional gas at the interface expects to be confirmed by experiments made by
using lateral contacts directly on the region between the $LMO$ and $SMO$ blocks. The d.c. conductivity
of the sheet could directly measure the density of carriers of the interface metal and confirm the
Drude-like low frequency behavior of in-plane response. Finally, one expects that a weak disorder present
in the system and not  included in our analysis can increase the scattering rate of the carriers
reducing the value of the in-plane conductivity for $\omega \rightarrow 0$.

\section{Conclusions}

In this paper we have discussed phase diagrams, spectral and
optical properties for a very large bilayer $(LMO)_{2n}/(SMO)_{n}$
(up to $48$ sites along the growth direction). A correlated
inhomogeneous mean-field approach has been developed in order to
analyze the effects of electron-lattice anti-adiabatic
fluctuations and strain. We have shown that a metallic ferromagnetic interface is
a quite robust feature of these systems for a large range of the electron-lattice couplings
and strain strengths. Furthermore, we have found that the size of the interface region
depends on the strength of electron-phonon interactions. At low temperature, the
general structure of our solutions is characterized by three
phases running along growth $z$-direction: antiferromagnetic phase
with localized/Delocalized charge carriers inside $LMO$ block, ferromagnetic
state with itinerant carriers at the interface, localized polaronic $G$-type
antiferromagnetic phase inside $SMO$ block. The type of
antiferromagnetic order inside $LMO$ depends on the strain induced
by the substrate.

Spectral and optical properties have been
discussed for different parameter regimes. Due to the formation of
the metallic interface, even in the intermediate to strong
electron-phonon coupling regime, the density of states never
vanishes at the chemical potential. Finally,  in-plane and
out-of-plane optical conductivities are sharply different: the
former shows a metallic behavior, the latter a transfer of
spectral weight at high frequency due to the effects of the
electrostatic potential well trapping electrons in $LMO$ block.
The in-plane response provides a signature of the formation of the
metallic ferromagnetic interface.

In this paper we have focused on static and dynamic properties at very low temperature.
The approach used in the paper is valid at any temperature. Therefore, it could be very interesting 
to analyze not only single interfaces, but also superlattices with different 
unit cells at finite temperature. Work in this direction is in progress.

\appendix
\section{}
In this Appendix we give some details about the effective
electronic Hamiltonian derived within our approach. After the
Hartree approximation for the long-range Coulomnb interactions,
the mean-field electronic Hamiltonian reads:

\begin{eqnarray}
H^{el}_{test}=-t\sum_{i_{||}}\sum_{i_{z}=1}^{N_{z}}\sum_{\delta_{||}}\gamma_{i_{z}}e^{-V_{i_{z}}}
c^{\dag}_{i_{||},i_{z}} c_{i_{||}+\delta_{||},i_{z}}
\nonumber \\
-t \sum_{i_{||}} \sum_{i_{z}=1}^{N_{z}} \sum_{\delta_{z}} \eta_{i_{z},i_{z}+\delta_{z}} e^{-W_{i_{z},i_{z}+\delta_{z}}}
c^{\dag}_{i_{||},i_{z}} c_{i_{||},i_{z}+\delta_{z}}
\nonumber \\
 + \sum_{i_{||}} \sum_{i_{z}=1}^{N_{z}} [ \phi(i_{z})-\mu ] c^{\dag}_{i_{||},i_{z}} c_{i_{||},i_{z}} +
N_{x}N_{y}(T_{1}+T_{2})
\nonumber \\
 +N_{x}N_{y}g^{2}\omega_{0}\sum_{i_z}\Delta_{i_z}+\sum_{i_{||}}\sum_{i_{z}=1}^{N_{z}}C_{i_z}(g^{2}\omega_{0})
c^{\dag}_{i_{||},i_{z}} c_{i_{||},i_{z}}.
\end{eqnarray}
The self-consistent Hartee potential is given by
\begin{eqnarray}
\phi(i_{z})=\frac{e^{2}}{\epsilon}[\sum_{j_{z}>i_{z}}\chi(j_{z})S(i_{z}-j_{z})+
\nonumber \\
\sum_{j_{z}<i_{z}}\chi(j_{z})S(i_{z}-j_{z})+S_{1}(0)\chi(i_{z})-S_{2}(i_{z})],
\end{eqnarray}
where the quantity $T_{1}$ is
\begin{eqnarray}
T_{1}=\frac{-e^{2}}{2\epsilon}[\sum_{i_{z}=1}^{N_{z}}\sum_{j_{z}>i_{z}}^{N_{z}}\chi_{i_{z}}\chi_{j_{z}}S(i_{z}-j_{z})+
\nonumber \\
\sum_{j_{z}<i_{z}}^{N_{z}}\chi_{i_{z}}\chi_{j_{z}}S(i_{z}-j_{z})+S_{1}(0)\sum_{i_{z}}^{N_{z}}\chi_{i_{z}}^{2}],
\end{eqnarray}
and $T_{2}$
\begin{eqnarray}
T_{2}=  \frac{e^{2}}{2\epsilon}[\sum_{I_{z}=1}^{N_{La}}\sum_{J_{z}>I_{z}}^{N_{La}}S(I_{z}-J_{z})+
\nonumber \\
\sum_{J_{z}<I_{z}}^{N_{La}}S(I_{z}-J_{z})+N_{La}S_{1}]
\end{eqnarray}
with $S(n_{z})$, $S_{1}(0)$ end $S_{2}(n_z)$
obtained by adding the Coulomb terms on in-plane lattice index.
The summations have been made modulating the Coulomb interaction with a
screening factor:
$\frac{e^{2}}{|\vec{r}_{i}-\vec{r}_{j}|} \rightarrow \frac{e^{2}e^{-\eta_S |\vec{r}_{i}-\vec{r}_{j}|}}{|\vec{r}_{i}-\vec{r}_{j}|}$,
where $\frac{1}{\eta_S}$ is a fictitious finite screening length in units of the lattice parameter $a$. Therefore,
$S(n_{z})$ is
\begin{equation}
S(n_{z})=\sum_{m_x,m_y} \frac{\exp{\left(-\eta_S \sqrt{m^2_x+m^2_y+n^2_z}\right)}}{\sqrt{m^2_x+m^2_y+n^2_z}},
\end{equation}
$S_{1}(0)$ is given by
\begin{equation}
S_{1}(0)=\sum_{m_x,m_y} \frac{\exp{\left(-\eta_S \sqrt{m^2_x+m^2_y}\right)}}{\sqrt{m^2_x+m^2_y}},
\end{equation}
with $(m_x,m_y) \neq (0,0)$, and $S_{2}(i_{z}-j_{z})$ is
\begin{equation}
S_2(n_{z})=\sum_{m_x,m_y} \sum_{i_z=1}^{l_z} \frac{ \exp{ \left(-\eta_S \sqrt{ h_x^2+h_y^2+h_z^2} \right)}}
{\sqrt{h_x^2+h_y^2+h_z^2} },
\end{equation}
with $l_z$ number of the planes of $LMO$ block, $h_x=m_x-0.5$, $h_y=m_y-0.5$, and $h_z=n_z-i_z-0.5$.

\end{document}